\documentclass{svproc}

\usepackage{cite}
\usepackage{url}
\usepackage{amsmath,amsfonts} 
\usepackage{float}
\usepackage{graphicx}
\usepackage{algorithm}
\usepackage{algpseudocode}
\usepackage{siunitx}
\usepackage[labelfont=bf]{caption}
\usepackage{subcaption}
\usepackage{tikz}
\usepackage{pgfplots}
\usepackage{epstopdf}
\usepackage{multirow}
\usepackage[font=scriptsize]{caption}
\usepackage{soul}
\usepackage{color, colortbl}
\usepackage{array}
\usepackage[T1]{fontenc}
\usepackage[utf8]{inputenc}
\usepackage{mathptmx}
\usepackage{listings}

\usepackage{enumitem}
\usepackage{hyperref}

\begin{document}
\mainmatter              

\title{Early Detection Of Mirai-Like IoT Bots In Large-Scale Networks Through Sub-Sampled Packet Traffic Analysis}
\titlerunning{Early Detection Of Mirai-Like IoT Bots}  
%
\author{Ayush Kumar \and Teng Joon Lim}	
\authorrunning{A. Kumar et al.} 
%
\tocauthor{Ayush Kumar and Teng Joon Lim}
\institute{Department of Electrical and Computer Engineering, \\ 
National University of Singapore, Singapore 119077 \\
			\email{ayush.kumar@u.nus.edu}, \email{eleltj@nus.edu.sg}}

\maketitle

\begin{abstract}
The widespread adoption of Internet of Things has led to many security issues. Recently, there have been malware attacks on IoT devices, the most prominent one being that of Mirai. IoT devices such as IP cameras, DVRs and routers were compromised by the Mirai malware and later large-scale DDoS attacks were propagated using those infected devices (bots) in October 2016. In this research, we develop a network-based algorithm which can be used to detect IoT bots infected by Mirai or similar malware in large-scale networks (e.g. ISP network). The algorithm particularly targets bots scanning the network for vulnerable devices since the typical scanning phase for botnets lasts for months and the bots can be detected much before they are involved in an actual attack. We analyze the unique signatures of the Mirai malware to identify its presence in an IoT device. The prospective deployment of our bot detection solution is discussed next along with the countermeasures which can be taken post detection. Further, to optimize the usage of computational resources, we use a two-dimensional (2D) packet sampling approach, wherein we sample the packets transmitted by IoT devices both across time and across the devices. Leveraging the Mirai signatures identified and the 2D packet sampling approach, a bot detection algorithm is proposed. Subsequently, we use testbed measurements and simulations to study the relationship between bot detection delays and the sampling frequencies for device packets. Finally, we derive insights from the obtained results and use them to design our proposed bot detection algorithm. 
\end{abstract}

\section{Introduction}
The Internet of things (IoT)\cite{iotsurvey} refers to the network of low-power, limited processing capability sensing devices which can  send/receive data to/from other devices using wireless technologies such as RFID (Radio Frequency Identification), Zigbee, WiFi, Bluetooth, 3G/4G etc. IoT devices are being deployed in a number of applications such as wearables, home automation, smart grids, environmental monitoring, infrastructure management, industrial automation, agricultural automation, healthcare and smart cities. Some of the popular platforms for IoT are Samsung SmartThings (consumer IoT for device management) and Amazon Web Services IoT, Microsoft Azure IoT, Google Cloud Platform (enterprise IoT for cloud storage and data analytics). The number of IoT devices deployed globally by 2020 is expected to be in the range of 20-30 billion \cite{iotforecast}. The number of devices has been increasing steadily (albeit at a slower rate than some earlier generous predictions), and this trend is expected to hold in the future.

IoT devices are being increasingly targeted by hackers using malware (malicious software) as they are easier to infect than conventional computers for the following reasons\cite{iotsecsurvey1,iotsecsurvey2,iotsecsurvey3}:
\begin{itemize}
\item There are many legacy IoT devices connected to the Internet with no security updates.
\item Security is given a low priority within the development cycle of IoT devices.
\item Implementing conventional cryptography in IoT devices is computationally expensive due to processing power and memory constraints.
\item Many IoT devices have weak login credentials either provided by the manufacturer or configured by users. 
\item IoT device manufacturers sometimes leave \textit{backdoors} (such as an open port) to provide support for the device remotely.
\item Often, consumer IoT devices are connected to the Internet without going through a firewall.  
\end{itemize}  
In a widely publicized attack, the IoT malware \textit{Mirai} was used to propagate the biggest DDoS (Distributed Denial-of-Service) attack on record on October 21, 2016. The attack targeted the Dyn DNS (Domain Name Service) servers \cite{miraiattack} and generated an attack throughput of the order of 1.2 Tbps. It disabled major internet services such as Amazon, Twitter and Netflix. The attackers had infected IoT devices such as IP cameras and DVR recorders with Mirai, thereby creating an army of bots (botnet) to take part in the DDoS attack. Apart from Mirai, there are other IoT malware which operate using a similar brute force technique of scanning random IP addresses for open ports and attempting to login using a built-in dictionary of commonly used credentials. BASHLITE \cite{bashlite}, Remaiten \cite{remaiten}, Hajime \cite{hajime} are some examples of these IoT malware. 
 
Bots compromised by Mirai or similar IoT malware can be used for DDoS attacks, phishing and spamming \cite{phishspam}. These attacks can cause network downtime for long periods which may lead to financial loss to network companies, and leak users' confidential data. McAfee reported in April 2017\cite{mcafee} that about 2.5 million IoT devices were infected by Mirai in late 2016. Bitdefender mentioned in its blog in September 2017\cite{bitdef} that researchers had estimated at least 100,000 devices infected by Mirai or similar malware revealed daily through telnet scanning telemetry data.  Further, many of the infected devices are expected to remain infected for a long time. Therefore, there is a substantial motivation for detecting these IoT bots and taking appropriate action against them so that they are unable to cause any further damage.

As pointed out in \cite{trillionflaws}, attempting to ensure that all IoT devices are secure-by-construction is futile as there will always be insecure devices (with patched and unpatched vulnerabilities) connected to the Internet due to the scale and diversity of IoT devices and vendors. Moreover, considering the lack of full-fledged operating systems, low power requirements, resource constraints and presence of legacy devices, it is practically unfeasible to deploy traditional host-based detection and prevention mechanisms such as antivirus, firewalls for IoT devices. Therefore, it becomes imperative that the security mechanisms for the IoT ecosystem are designed to be network-based rather than host-based.  

In this research, we propose a network-based algorithm which can be used to detect IoT bots infected by Mirai-like malware (which use port-based scanning) in large-scale networks. Bots scanning the network for vulnerable devices are targeted in particular by our algorithm. This is because the scanning and propagation phase of the botnet life-cycle stretches over many months and we can detect and isolate the bots before they can participate in an actual attack such as DDoS. If the DDoS attack has already occurred (due to a botnet), detecting the attack itself is not that difficult and there are already existing methods both in literature and industry to defend against such attacks. Moreover, our algorithm is practical in terms of utilization of computational resources (such as CPU processing power, memory). For example, ISP (Internet Service Provider) network operators can use the proposed algorithm to identify infected IoT devices in their network. The operators can then take suitable countermeasures such as blocking the traffic originating from IoT bots and notifying the local network administrators. Actions that can be taken post bot detection are further discussed in a later section. 
The major contributions of this paper are listed below:
\begin{enumerate}
\item We have analyzed the traffic signatures produced by Mirai malware infecting IoT devices through testbed experiments. Further, we have identified specific signatures which can be used to positively detect the presence of Mirai and similar malware in IoT devices. These signatures are similar to the observations reported in \cite{miraiusenix} based on their analysis of the Mirai source code.
\item We have proposed an algorithm to detect Mirai-like IoT malware bots in large-scale networks. The algorithm is based on a novel two dimensional sampling approach where the device packets are sampled across time as well as across the devices.
\end{enumerate}

The rest of the contents of this paper are organized as follows. In Section \ref{literature}, we review few prominent works on detecting botnets exploiting CnC communication features and intrusion detection systems for IoT. Subsequently, in section \ref{miraitrafficanalysis}, we explain the operation of Mirai, extract important features from the traffic generated by Mirai bots in a testbed and present a detailed analysis of those features towards detecting Mirai-like bots. Section \ref{botdetection} presents the network deployment of our bot detection solution. It also includes the formulation of the optimization problem resulting from detection of IoT bots along with the constraints imposed by limited computational resources followed by the proposed bot detection algorithm. Finally, the algorithm is numerically evaluated and the results are presented in section \ref{eval}.

\section{Related Work}
\label{literature}
There are several works in the literature on detecting botnets using their CnC communication features. We list a few prominent ones in this section. The authors in \cite{ircbotnet} present machine-learning based classification methods to detect CnC traffic of IRC (Internet Relay Chat) botnets by differentiating between IRC and non-IRC traffic and then differentiating between bot and real IRC traffic. 
Bothunter\cite{bothunter} builds a \textit{bot infection dialog model} using the network communication flows between internal hosts and external entities during successful bot infections. Three bot-specific sensors are constructed based on the dialog model and correlation is performed between inbound intrusion/scan alarms and the infection dialog model to generate a consolidated report. Spatio-temporal similarities between bots in a botnet	 in terms of bot-CnC coordinated activities are captured from network traffic and leveraged towards botnet detection in a local area network in Botsniffer\cite{botsniffer}. In BotMiner\cite{botminer}, the authors have proposed a botnet detection system which clusters similar CnC communication traffic and similar malicious activity traffic, and uses cross cluster correlation to detect bots in a monitored network. It is claimed to be independent of CnC protocol and structure with no requirement of \textit{a priori} knowledge about the botnets. 
A system for detecting covert P2P (Peer-to-Peer) botnets has been proposed in \cite{p2pbotnet}. After extracting the statistical CnC communication features for P2P botnets, the botnet detection system utilizes them to distinguish between legitimate and malicious P2P traffic.

There has also been some research on intrusion detection and anomaly detection systems for IoT. A whitelist-based intrusion detection system for IoT devices (Heimdall) has been presented in \cite{heimdall}. Heimdall is based on dynamic profile learning and is designed to work on routers acting as gateways for IoT devices. The authors in \cite{tldrtc} propose an intrusion detection model for IoT backbone networks leveraging two-layer dimension reduction and two-tier classification techniques to detect U2R (User-to-Root) and R2L (Remote-to-Local) attacks. In a recently published paper \cite{nbaiot}, deep-autoencoders based anomaly detection has been used to detect attacks launched from IoT botnets. The method consists of extraction of statistical features from behavioral snapshots of normal IoT device traffic captures, training of a deep learning-based autoencoder (for each IoT device) on the extracted features and comparison of the reconstruction error for traffic observations with a threshold for normal-anomalous classification. The proposed detection method was evaluated on Mirai and BASHLITE botnets formed using commercial IoT devices.

While a number of above anomaly detection works leverage ML (machine learning)-based approaches, there are several issues associated with them\cite{MLIDissues}. One of the major issues is the occurrence of false positives. Even a small percentage of false positives, e.g. 1\% which is considered acceptable in academic research on anomaly detection, can lead to thousands of alerts per day based on the traffic volume processed\cite{FArates}. Both false positives and false negatives have costs (e.g. financial expenses for an organization) associated with them, with the cost associated with false negatives typically being much higher. Second, many research works on anomaly detection using ML fail to explain why a particular ML algorithm would perform well in the system under consideration.  Third, many ML algorithms are suitable for offline batch operations rather than low-latency real-time detection. Finally, instead of starting with the premise of using ML approach for a detection task which is a common flaw in anomaly detection research, one should carry out a neutral evaluation of all the available tools for the task and then decide on the most appropriate one.  

Our work addresses a few important gaps in the literature when it comes to distinguishing between legitimate and botnet IoT traffic. First, almost all the works cited above on detecting botnets using their CnC communication features \cite{ircbotnet, bothunter,botsniffer,botminer} utilize all the packets transmitted by all the devices in a monitored network for a specific time period towards designing a botnet detection solution. This approach is highly impractical if the resulting solution is to be deployed for IoT devices in real world networks. The reason is that processing all the packets for all devices in a large network would require a lot of computational resources. Second, our focus is not only on detecting bots employing IRC-based CnC communications as done in \cite{ircbotnet}. The bot detection algorithm proposed by us in Section \ref{optprob} is independent of the bot-CnC communication protocol. Third, we do not aim to detect botnets (networks of bots) but instead, individual bots. Therefore, we don't require computationally expensive clustering algorithms as used in \cite{botsniffer,botminer}. 

Fourth, we do not extract CnC communication features and use them to identify bot-CnC communications as done in \cite{botsniffer,botminer,p2pbotnet}. This is because we aim to detect bots infected by Mirai-like IoT malware, towards which much simpler features can be used as discussed in Section \ref{miraitrafficfeatures}. Fifth, unlike \cite{nbaiot}, we aim to detect IoT bots much before the actual attack, during the scanning phase itself as explained in Section \ref{botdetection}. Finally, most of the above cited works use quantifiers such as detection rate and false positive rates to evaluate the performance of their proposed botnet detection solutions. Instead, we use a quantity called average detection delay (defined in Section \ref{optprob}) for the performance evaluation of our proposed bot detection solution since the features used by our solution eliminate the possibility of inaccurate detections or false positives. 
To the best of our knowledge, there are no existing papers on detecting IoT bots compromised by Mirai or its variants which exhibit TELNET port-based scanning behavior. 
  

\section{Mirai Traffic Analysis}
\label{miraitrafficanalysis}
Detecting IoT devices compromised by Mirai-like malware requires us to analyze the packet traffic generated by those devices and extract some features to aid us in detection. 
In this section, we begin with a brief description the operation of Mirai to make the readers familiar with some of the related terms. Later, we present a testbed that we use to emulate IoT devices, infect them with Mirai and capture the packet traffic generated from them. Finally, we present the extracted features and analyze them in detail with respect to identifying Mirai bots.

\subsection{Mirai Operation}
\begin{figure}[h]
\centering
\includegraphics[scale=0.4]{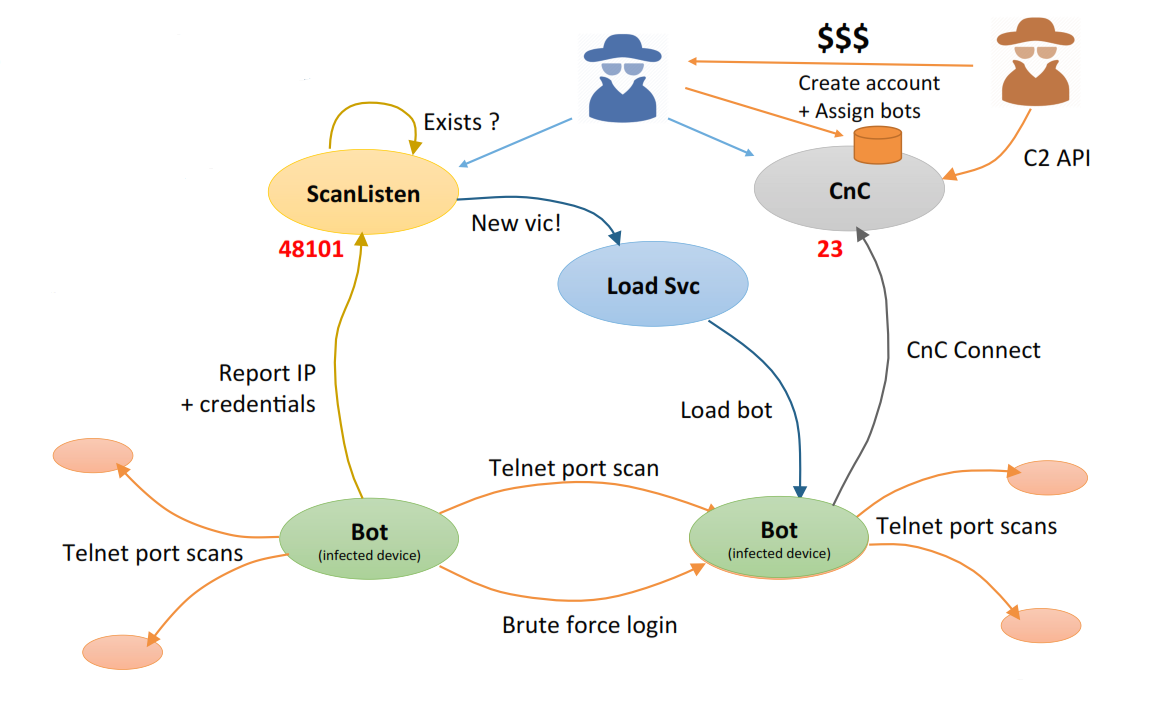}
\caption{Operation of various components of Mirai \textit{(Source: Radware \cite{miraipres})}}
\label{mirai_op}
\end{figure}

The Mirai \cite{mirai} setup consists of three major components: \textit{bot}, \textit{scanListen}/\textit{loading} server, and the \textit{CnC} (Command-and-Control) server. The \textit{CnC} server also functions as a MySQL\cite{mysql} database server. User accounts can be created in this database for customers who wish to hire DDoS-as-a-service.
The operation of Mirai is illustrated in Fig. \ref{mirai_op}. Once an IoT device is infected with Mirai (and becomes a bot), it first attempts to connect to the listening \textit{CnC} server by resolving its domain name and opening a socket connection. Thereafter, it starts scanning the network by sending SYN packets to random IP addresses and waiting for them to respond. This process may take a long time since the bot has to go through a large number of IP addresses. Once it finds a vulnerable device with a TELNET port open, it attempts to open a socket connection to that device and emulates the TELNET protocol. Then it attempts to login using a list of default credentials and if working credential is found, it reports the IP address of the discovered device and the working TELNET login credentials to the listening \textit{scanListen} server.
The \textit{scanListen} server sends that information to the \textit{loader} which again logs in to the discovered device using the details received from the \textit{scanListen} server. Once logged in, the \textit{loader} downloads the Mirai bot binary to that device and the new bot connects to the \textit{CnC} server and starts scanning the network.

\subsection{Testbed Description}
\begin{figure}[h]
\centering
\includegraphics[scale=0.4]{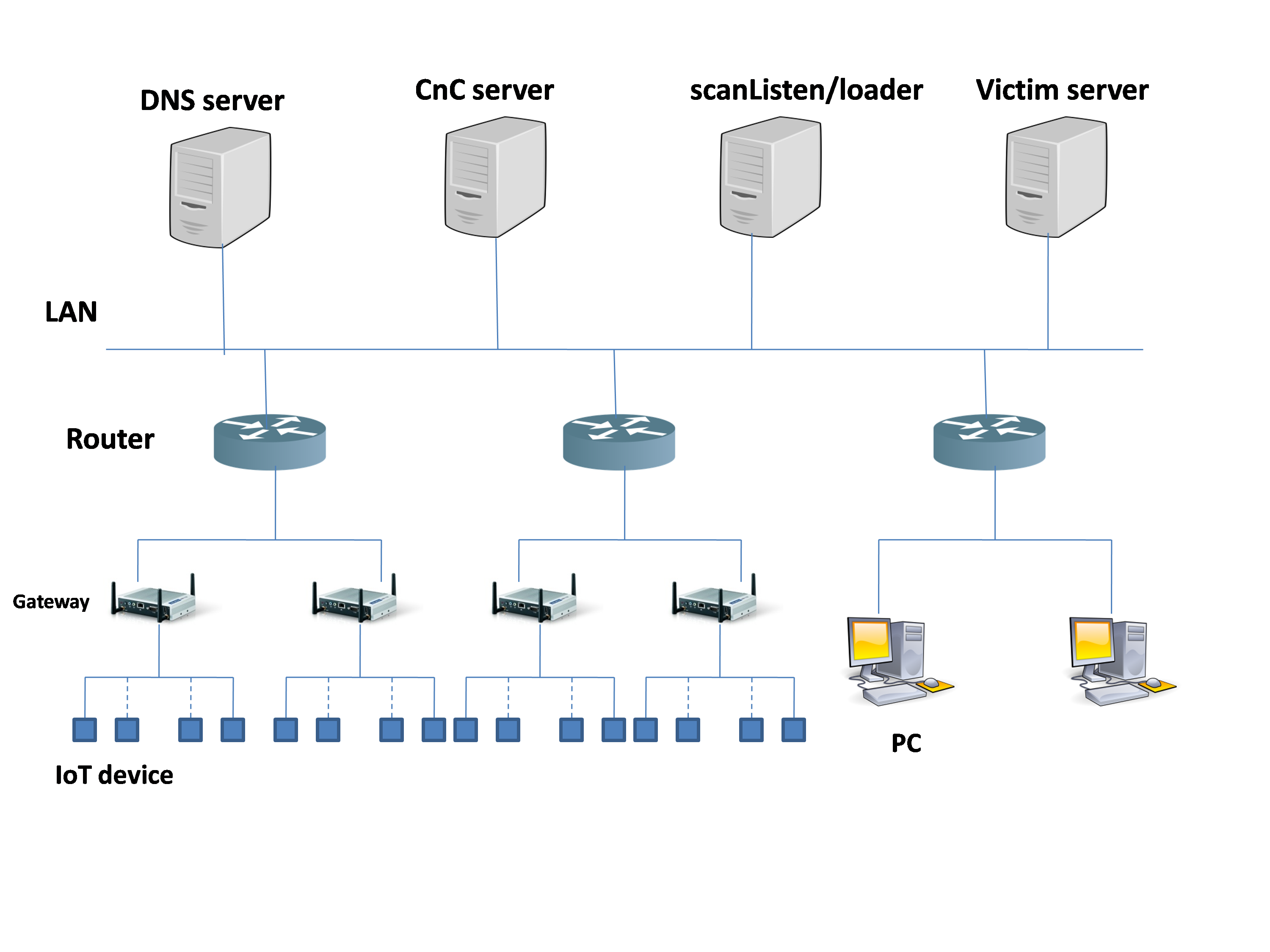}
\caption{Testbed used to emulate Mirai behavior}
\label{mirai_testbed}
\end{figure}

The testbed shown in Fig. \ref{mirai_testbed} was configured on an isolated computing cluster. Each cluster node has two Intel Xeon E5-2620 processors, 64 GB DDR4 ECC memeory and runs Ubuntu 14.04 LTS standard image. The testbed consists of a local authoritative DNS server, a CnC (Command-and-Control) server and a server for scanListen and loading utility, all connected to a single LAN. The IoT gateways are connected to the above LAN through routers and behind the gateways are QEMU\cite{qemu}-emulated IoT devices (Raspberry Pi). We chose this gateway-IoT device topology since it is used in a number of IoT deployments (such as IP cameras, smart lighting devices, wearables etc.). The testbed also includes few non-IoT devices (PCs) to reflect real-world networks. As per our information, this is the first controlled testbed to emulate the true behavior of Mirai malware. It can be modified to add more nodes, study a different network topology and test more advanced versions or derivatives of Mirai malware.

\subsection{Mirai Traffic Features}
\label{miraitrafficfeatures}
We infected the emulated IoT devices in our testbed with Mirai and captured a total of 1,583,623 packets transmitted by the devices. An analysis of the captured packets reveals the following features/signatures:
\begin{itemize}
\item The scanning packets are all TCP SYN (synchronization) packets.
\item The destination port numbers of scanning packets are distributed as $\sim$90\% port 23 and $\sim$10\% port 2323. No other port numbers are observed.
\item There is a periodic exchange of keep alive messages (PSH+ACK) between the bot and the CnC server. PSH refers to a push message and ACK refers to acknowledgement.
\end{itemize}
Both ports 23 and 2323 are assigned for TELNET applications\cite{ianaportreg,port2323}.
The TELNET\cite{telnetrfc} protocol is used for bidirectional byte-oriented communication. In the most widely used implementation of TELNET, a user with a terminal and running a TELNET client program, accesses a remote host running a TELNET server by requesting a connection to the remote host and logging in by providing its credentials. The most common application of TELNET is for configuring network devices such as routers. Now, IoT devices operate by continuously transmitting sensed data to and receiving commands from cloud servers through a gateway over a secure communication channel without external human input\cite{googleiot}. We claim that an IoT device is unlikely to be used to access or configure another device using TELNET, and therefore in the absence of malware infection, IoT devices should not open TELNET connections to any other device.

To verify our claim that uninfected IoT devices are not expected to open TELNET connections, the following experiment was conducted. We configured a Raspberry Pi 3 (Model B+) to act as a gateway and connected it to several real-world IoT devices such as IP cameras (D-Link), motion sensors (D-Link), smart bulbs (Philips Hue), smart switches (WeMo) and smart plugs (TPLink). We left the devices connected for a long time and for each device type mentioned above, we captured around 10,000 packets per device at the gateway interface. Later, the captured packets were analysed using Wireshark \cite{wireshark} and no SYN packets with destination ports 23 or 2323 were found. 
Thus, if a SYN packet from an IoT device with destination port number 23 or 2323 is received, it is sufficient evidence to conclude with certainty that the IoT device is infected with a Mirai-like malware. The above experiment also help us to rule out \textit{false positives}, if any at all, if we use the identified scanning traffic signatures, which is a substantial advantage when it comes to practical intrusion detection.

The third Mirai signature related to keep-alive messages is not required since the port-scanning signatures is sufficient for detection with certainty. We may require the third signature to detect more advanced malware which do not use TELNET port-based scanning. It needs to be emphasized here that the TELNET port-scanning signatures can be used to identify not only bots infected by Mirai but also other Mirai-like malware such as BASHLITE, Remaiten, Hajime etc. which employ similar TELNET port brute forcing technique.

\section{Mirai-like IoT Malware Bot Detection}
\label{botdetection}
The bot scanning traffic analyzed in the previous section can be detected using simple firewalls. However, since IoT devices are usually resource-constrained, they do not have firewalls installed on them. Moreover, network-level firewalls (protecting computers in a LAN/WAN/intranet) are usually not configured to block TELNET traffic. 
In this section, we discuss the network deployment of our bot detection solution. Further, we formulate the optimization problem arising out of detecting IoT bots with the accompanying computational resource constraints. Further, we propose an algorithm for bot detection based on our analysis.

\subsection{Bot Detection Solution Deployment}
\label{implement}
\begin{figure}[h]
\centering
\includegraphics[scale=0.4]{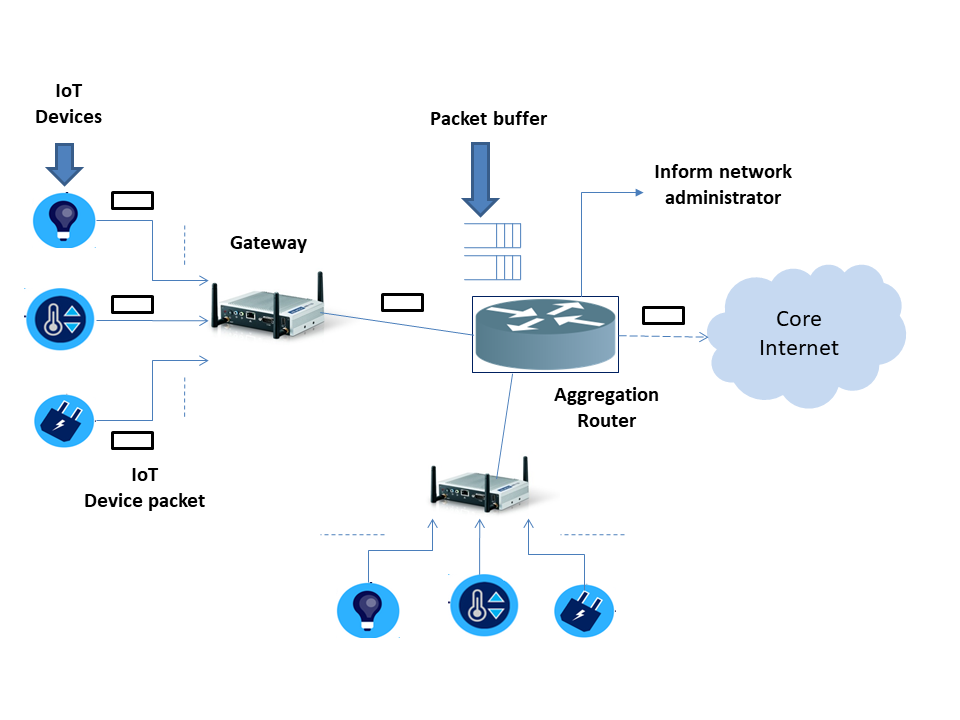}
\caption{Prospective network deployment for proposed bot detection solution}
\label{sentinel_arch3}
\end{figure}

It is proposed that our Mirai-like bot detection solution be run on the edge gateway connected to IoT devices or the aggregation router connected directly to the gateway.  Assuming that we run the bot detection solution on the router, a prospective network deployment for our solution is shown in Fig. \ref{sentinel_arch3}. The incoming packets at the router are arranged and stored in buffers according to their source devices. 

We process only IoT device packets at the router, whereas the aggregate router traffic consists of IoT as well as non-IoT traffic (PCs, smartphones etc.) 
The authors in \cite{profiliot} distinguish between traffic generated by IoT and non-IoT devices from a single TCP session by analyzing user-agent HTTP property for smartphones and single-session binary classifiers for PCs. A classification accuracy of 100\% for smartphones and false positive, negative rates of 0.003 each for PCs were claimed to be achieved. We can use their methods to distinguish between IoT and non-IoT device packets using a single session worth of packets. Further, once we identify a device as belonging to IoT or non-IoT type, we can continue to use this information in the future as the device type is not expected to change. 

It is assumed that the ISPs already have access to the information regarding vulnerable and non-vulnerable devices. As explained earlier in Section \ref{optprob}, IoT devices installed in home environments can be regarded as vulnerable while the devices installed in enterprise/industrial/government networks can be deemed as non-vulnerable.
We expect the firmware running on bot detection routers to be upgradeable so that in future, if more advanced bot detection algorithms are designed (e.g. for IoT malware which do not just rely on port based scanning), the corresponding software updates can be easily pushed to those routers. 

Once the bots are detected by our proposed algorithm, the next step is to take mitigating actions to prevent the bots from spreading further damage. The network administrator can block the entire traffic originating from bots and bring them back online only after it is confirmed that the malware has been removed from those IoT devices. The concerned ISP can inform the device owners and ask them to secure their device (by using strong usernames/passwords, placing the device behind a firewall etc.). Another defense mechanism is that instead of blocking all the traffic, the bot can be allowed communications with a few secure domains for remediation of malware infection. This strategy has been mentioned as part of the bot remediation techniques\cite{ietfbot} recommended for ISPs by IETF (Internet Engineering Task Force). The bot can also be  placed under continuous monitoring and all other communication except that required for the underlying IoT device to function can be denied. Finally, security personnel can exploit bugs in the bot binary to disinfect them remotely.

\subsection{Formulation of Optimization Problem}
\label{optprob}
Processing all the incoming packets at the bot detection router to check if they originated from an IoT device and subsequently matching those packets against the Mirai traffic signature would require a lot of memory. To give an example, considering 10,000 IoT devices connected to the aggregation router with each device transmitting at 10\% of the peak data rate of \textit{250 kbits/s} (according to IEEE 802.15.4 standard for low rate personal wireless devices), we need $\approx$ 9.36GB of storage for a 5 mins traffic session. However, typically WAN aggregation routers have 1-4GB RAM and few GBs of external storage only, of which a major part is used in packet routing. Therefore, for our bot detection solution, we propose to sample only a fraction of the IoT devices per unit time. However, this approach has the drawback that we may miss the scanning packets due to the sub-sampling operation. This leads to the formulation of the following optimization problem to detect infected devices.

Our objective in this optimization problem is to minimize the cost associated with the delay in detecting a compromised device. We define \textit{average detection delay} ($T_D$) as the average time between the first occurrence of a scanning packet and the positive conclusion that the originating device is infected. Now, some IoT devices in a network are easier to infect with malware than others. Therefore, we split the IoT devices into two categories: \textit{vulnerable} and \textit{non-vulnerable} devices. Vulnerable devices are the devices which are easier to get successfully infected with Mirai-like malware and added to the botnet. The devices other than vulnerable ones are non-vulnerable devices. 

For example, personal IoT devices installed at homes can be deemed as vulnerable since they are less likely to be behind a firewall (host-level firewalls not feasible on IoT devices due to resource constraints) and more likely to have their TELNET ports open (often owners buy cheap devices in which the manufacturer has left TELNET port open for remote configuration etc.). IoT devices installed in enterprise/industrial/government networks can be categorized as non-vulnerable since most likely, they would be behind a network-level firewall (blocking access to insecure TELNET connections) and they are much less likely to have to have their TELNET ports open (due to organizational IT security policies). 


We define the \textit{sampling frequency for an IoT device} as the fraction of the time when that device is selected for monitoring for possible infection. We also define the \textit{sampling matrix}, $\Sigma$ as a matrix with columns representing devices and rows representing the packets transmitted by those devices. An element of $\Sigma$ is equal to $1$ when the corresponding packet has been sampled and equal to $0$ when the corresponding packet has not been sampled.  

Further, our optimization problem imposes the following constraints that need to be satisfied:
\begin{itemize}
\item The sampling frequency for a vulnerable device ($f_n^v$) should be greater than the sampling frequency for a non-vulnerable device ($f_n^{nv}$). This is because vulnerable devices are more likely to be attacked than non-vulnerable devices and hence they need to be more frequently monitored.  
\item The total number of vulnerable and non-vulnerable devices selected within a certain time period ($\rho_v f_n^v T+\rho_{nv} f_n^{nv}T$) should not exceed a maximum number ($f_n^{max}T$), where $\rho_v$ and $\rho_{nv}$ are the fractions of total number of devices that and vulnerable and non-vulnerable respectively. This is to limit the utilization of computational resources for if the total number of selected devices is more than an upper bound, it may require significant amounts of processing power defeating the purpose of packet sub-sampling.
\item The maximum number of vulnerable devices selected at any time should have an upper bound ($N_v^{max}$). Similarly, the maximum number of non-vulnerable devices selected at any time should have an upper bound ($N_{nv}^{max}$). This is again to place a bound on computational resources utilization.  
\item After a certain number of sampling time units ($T$), every device (in the set of all devices, $\Omega_N$) should be covered by the sampling process. This is to ensure that every device is checked for malware infection within a certain time duration or else few devices which are infected may be missed by the sampling process.
\end{itemize}
We propose to minimize the cost associated with the average detection delay while satisfying the above constraints as follows:
\begin{equation*}
\begin{aligned}
& \underset{\Sigma, f_n^v, f_n^{nv}}{\text{minimize}}
& & \alpha T_D (f_n^v, f_n^{nv}, Y_v, Y_{nv}) \\
& \text{subject to}
& & f_n^v > f_n^{nv} \\
&&& \rho_v f_n^v+\rho_{nv} f_n^{nv}<f_n^{max} \\
&&& max [N_v\{\Sigma\}]<N_v^{max} \\ 
&&& max [N_{nv}\{\Sigma\}]<N_{nv}^{max} \\
&&& \bigcup_{t=t_{start}}^{t_{start}+T} dev\_set(\Sigma, t)=\Omega_N
\end{aligned}
\end{equation*}

where $\alpha$ is defined as the cost incurred by the bot detection algorithm due to a unit average detection delay, $N_v\{.\}, N_{nv}\{.\}$ denote the number of vulnerable and non-vulnerable devices selected in $\Sigma$ at any point of time, $Y_v$ is the set of vulnerable devices, $Y_{nv}$ is the set of non-vulnerable devices, and $dev\_set(.)$ is a function that outputs the set of devices sampled in $\Sigma$ at a time $t$. It is to be noted that the above optimization problem is a combinatorial one and it is computationally hard to find an optimal solution\cite{combincomplex}. Hence, we devise a method to numerically solve the optimization problem. The results obtained from the numerical analysis are explained in Section \ref{eval}. Based on our findings through the formulation of optimization problem, we have proposed an algorithm for detecting IoT bots (shown in Algorithm \ref{A1}) which is practical in terms of lower number of packets that need to be monitored for infected device detection. The values for $f_n^v$ and $f_n^{nv}$ to be used while designing our algorithm will be discussed in our numerical analysis. 
\begin{algorithm}[h]
	\centering	
	\caption{\scriptsize{IoT Bot Detection Algorithm}}
	\label{A1}
	\begin{algorithmic}[1]
		\State Initialize $\Sigma$, NUM\_PKTS, t.
		\For {$pktcnt=1$ to NUM\_PKTS}
		\If {src\_dev(recv\_pkt) $\not\in$ list\_dev}
		\State add\_dev\_to\_list(src\_dev(recv\_pkt),list\_dev)
		\EndIf
		\State add\_pkt\_to\_buf(recv\_pkt, dev\_buf(src\_dev(recv\_pkt))
		\State pktcnt$=$pktcnt+1
		\EndFor
	    \While {TRUE}
	    \State sel\_dev\_set$=$dev\_set($\Sigma$,t)
	    \For {$i=1$ to length(sel\_dev\_set)}
	    \State sampled\_pkts(t,:)=dev\_buf(sel\_dev\_set(i), CURRENT\_PKT)	
	    \EndFor
	    \For {$j=1$ to length(sampled\_pkts(t,:))}
		\If {Check\_TCP\_flag(sampled\_pkts(t,j)) $=$ SYN \& Check\_dst\_port(sampled\_pkts(t,j)) $=$ 23 OR 2323}	
		\State Bot\_detected(src\_dev(sampled\_pkts(t,j))) $=$ TRUE
		\EndIf
		\EndFor
		\State t$=$t+1
		\EndWhile
	\end{algorithmic}
\end{algorithm}

\section{Evaluation of Proposed Algorithm}
\label{eval}
\begin{figure}[h]
\centering
\includegraphics[scale=0.4]{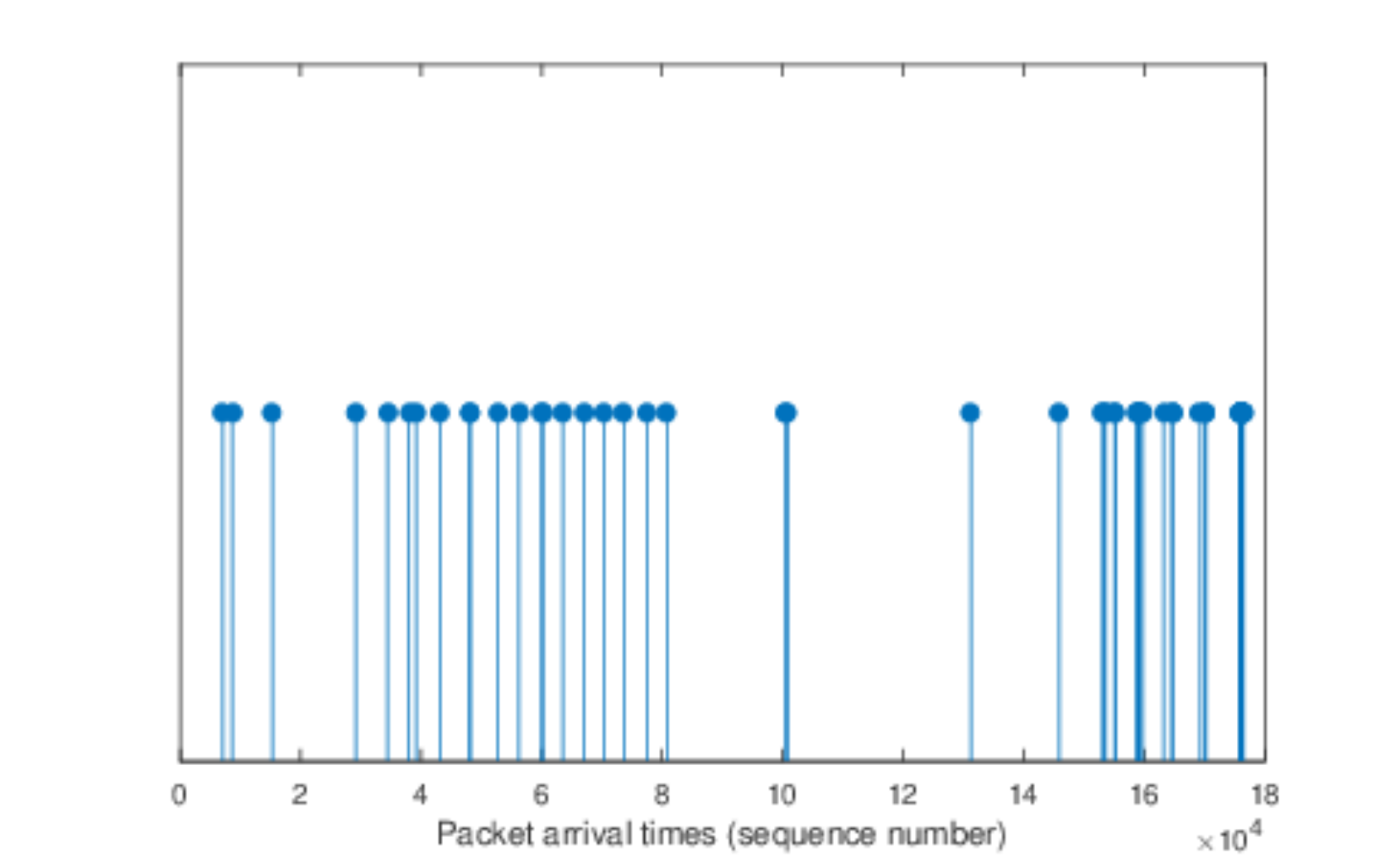}
\caption{Arrival times of scanning packets}
\label{scanpkt_arrivals2}
\end{figure}

In this section, we analyze the the behavior of average detection delay for vulnerable and non-vulnerable devices with varying sampling rates. 
A few important background details are presented below:
\begin{itemize}
\item The set of attacked devices, $\Phi$ is selected based on the assumed probability model for malware attack on vulnerable and non-vulnerable devices. For example, we can assume the probability of attack on vulnerable devices within a given time duration ($N_p$ packets' transmission) as $p_1$ and that on non-vulnerable devices as $p_2$.
\item The sampling matrix, $\Sigma$ used in our evaluation has a staggered structure and may be visualized as in Fig. \ref{samp_mat}. Since the sampling frequency for vulnerable devices is greater than that for non-vulnerable devices, the portion of $\Sigma$ containing packets transmitted by vulnerable devices has a more dense distribution of $1$s than that for non-vulnerable devices. The structure of the matrix also ensures that every device is sampled after a certain number of sampling time units as required by one of the constraints in the optimization problem presented in Section \ref{optprob}.
\begin{figure}[h]
\centering
\includegraphics[scale=0.6]{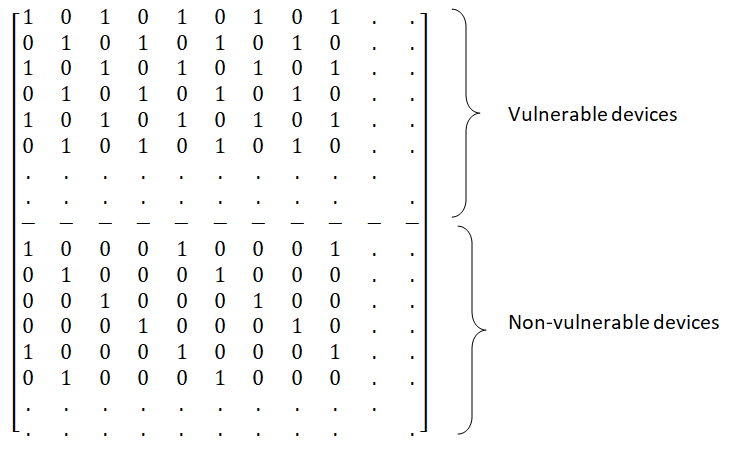}
\caption{Sampling matrix example}
\label{samp_mat}
\end{figure} 

\item We form a \textit{scanning} matrix with size as (number of IoT devices) $\times$ (number of packets transmitted). The matrix uses $0$ to represent a normal IoT device packet and $1$ to represent a malware scanning packet. Only the devices in $\Phi$ would have $1$s in their corresponding rows in the scanning matrix.
\item The elements where the scanning and the sampling matrices are both $1$ represent detected scanning packets. This is because the matching elements would only be present where the scanning packet transmitted by an attacked device has been selected by the sampling process. 
\end{itemize}
Moreover, we need to form a statistical model for scanning packet arrivals in the scanning matrix. Towards this, we used one of our emulated IoT devices and established a video streaming server to emulate the operation of an IP camera (IoT device used in Mirai attack on Dyn). Another emulated IoT device acted as a client connected to the video stream. The other emulated devices were configured to have their TELNET port number 23 open and listening for connections. Subsequently, we infected the video streaming device with Mirai and captured the transmitted packets at its gateway interface using Wireshark. Our observations from the packet capture are listed below: 
\begin{itemize}
\item The video streaming packets are transmitted almost continuously. The transmission is interrupted only by bot-CNC server communication packets, scanning packets and some other types of packets such as ARP (Address Resolution Protocol).
\item The bot scanning packets are sometimes transmitted within short intervals and at other times they are transmitted far apart as shown in Fig \ref{scanpkt_arrivals2}.
\end{itemize}  
Based on the above empirical observations, we model the scanning packet arrivals as a Poisson process, i.e., the inter-packet arrival times for scanning packets are exponentially distributed with the average packet arrival rate calculated from the testbed measurements. At all other times, we assume that normal IoT traffic is transmitted, again based on above observations.

\begin{figure}[h]
\centering
\includegraphics[scale=0.4]{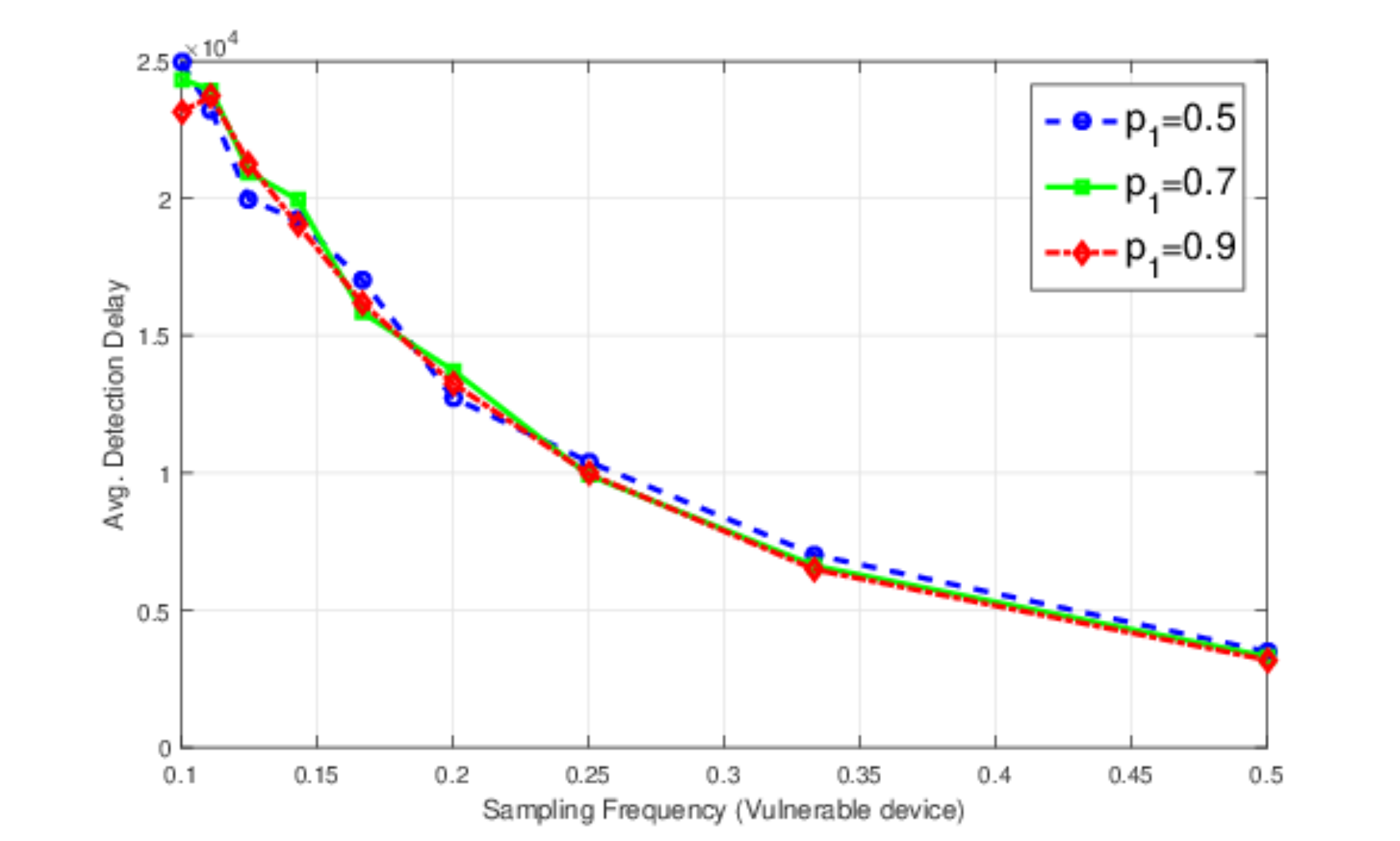}
\caption{Average detection delay vs sampling frequency plot for vulnerable devices}
\label{sim1}
\end{figure}

\begin{table}
	\centering
	\caption{Parameter values assumed in numerical analysis}
	\label{paramtable}
    \begin{tabular}{ | l | l |}
    \hline
    \textbf{Parameter} & \textbf{Value} \\ \hline
    $N_v^{max}$ & 40 \\ \hline
    $N_{nv}^{max}$ & 80 \\ \hline
    $f_n^{max}$ & 0.5 \\ \hline
    $N_p$ & 50 \\ \hline
    Total no. of IoT devices & 100 \\ \hline
    \% age of vulnerable devices & 40 \\ \hline
    No. of packets transmitted per device & 100,000 \\ \hline
    Avg. rate of arrival of scanning packets & 3386 \\
    (per packet elapsed) & \\
    \hline
    \end{tabular}
\end{table}
The values assumed for the various parameters in our analysis are shown in Table \ref{paramtable}.
The plot for average detection delay vs sampling frequency for different values of attack probability on vulnerable devices ($p_1$) is shown in Fig. \ref{sim1}. The detection delay values are averaged over all the detected devices as well as over a number of trial runs (1000). The units of average detection delay are in \textit{number of packets elapsed} while the units of sampling frequency are in \textit{per packet elapsed}. It can be observed that the average detection delay decreases almost exponentially with increasing sampling frequency. This behavior can be intuitively explained as follows. Increasing the sampling frequency means that the vulnerable devices are sampled much more frequently, which in turn increases the likelihood of sampling the scanning packets transmitted by infected vulnerable devices. Once a scanning packet is sampled, it can be positively concluded that the corresponding source device is infected as discussed in section \ref{miraitrafficfeatures}. Hence, an increase in the likelihood of sampling scanning packets should lead to a decrease in the average detection delay as defined in section \ref{optprob}. Further, it can also be noted from the plot that increasing the sampling frequency beyond a certain value (e.g. '0.33' for $p_1=0.5$) leads to slower reduction in average detection delay. This suggests that while designing the proposed Algorithm \ref{A1}, the sampling frequency for vulnerable devices should be selected towards the upper half of the range of available values but not too high since higher sampling frequencies will not result in more benefit in terms of decrease in average detection delay. Instead, sampling frequencies which are too high may lead to greater consumption of computational resources. 

One may observe that the average detection delay values decrease slightly as the attack probability increases. This is expected since an increase in attack probability means that more number of vulnerable devices are likely to be infected, thus increasing the likelihood of sampling the scanning packets transmitted by those infected devices resulting in a decrease in average detection delay. Lastly, the plots for the three attack probabilities, $p_1=0.5, 0.7, 0.9$, are quite close to each other, suggesting that changes in attack probability do not affect the average detection delay vs sampling frequency behavior significantly. 

\begin{figure}[h]
\centering
\includegraphics[scale=0.4]{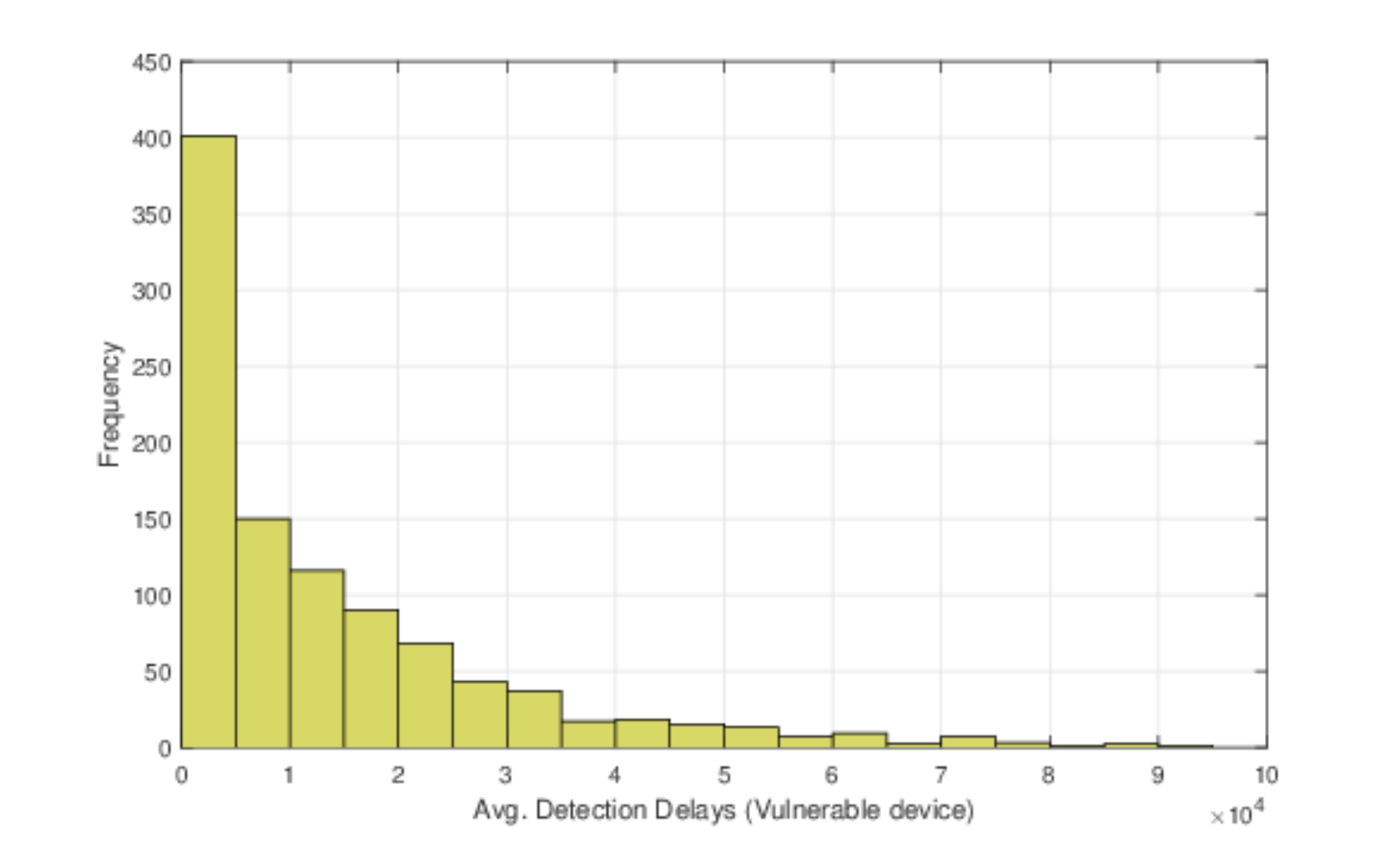}
\caption{Histogram of Average detection delays for vulnerable devices}
\label{sim2}
\end{figure}

In Fig. \ref{sim2}, we have illustrated the distribution of average detection delays for vulnerable devices for a sampling frequency of 0.2 and attack probability of 0.6 using a histogram. The distribution closely fits an exponential distribution with a mean of $\approx 52$, suggesting that the probability of achieving higher and higher average detection delays for vulnerable devices decreases almost exponentially. Vulnerable devices are sampled at a relatively higher frequency and also have a higher probability of being infected than non-vulnerable devices. Therefore, scanning packets can be detected with lower delays in most trials, resulting in higher probability for lower values and lower probabilities for higher values of average detection delays.

\begin{figure}[h]
\centering
\includegraphics[scale=0.4]{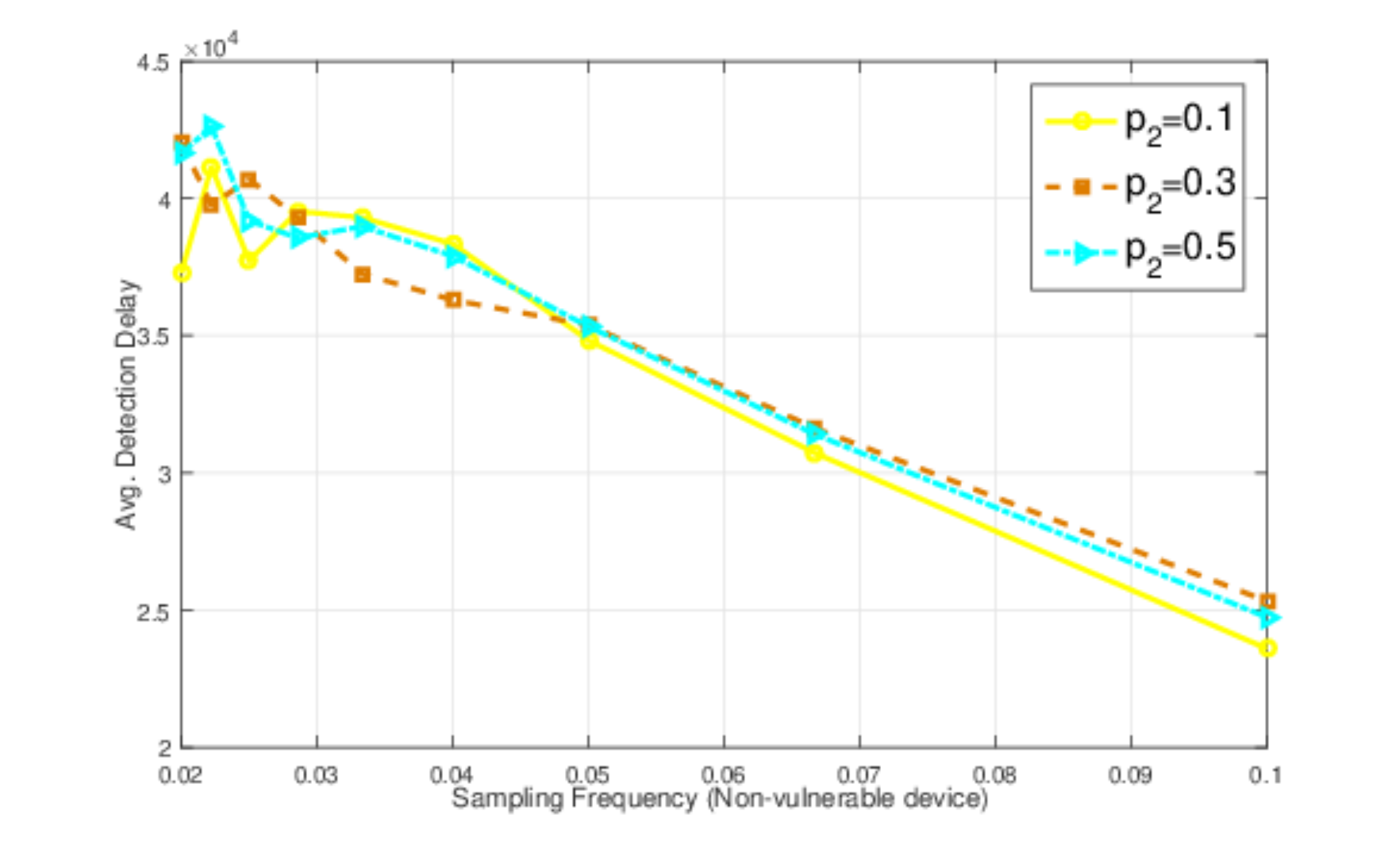}
\caption{Average detection delay vs sampling frequency plot for non-vulnerable devices}
\label{sim3}
\end{figure}

\begin{figure}[h]
\centering
\includegraphics[scale=0.4]{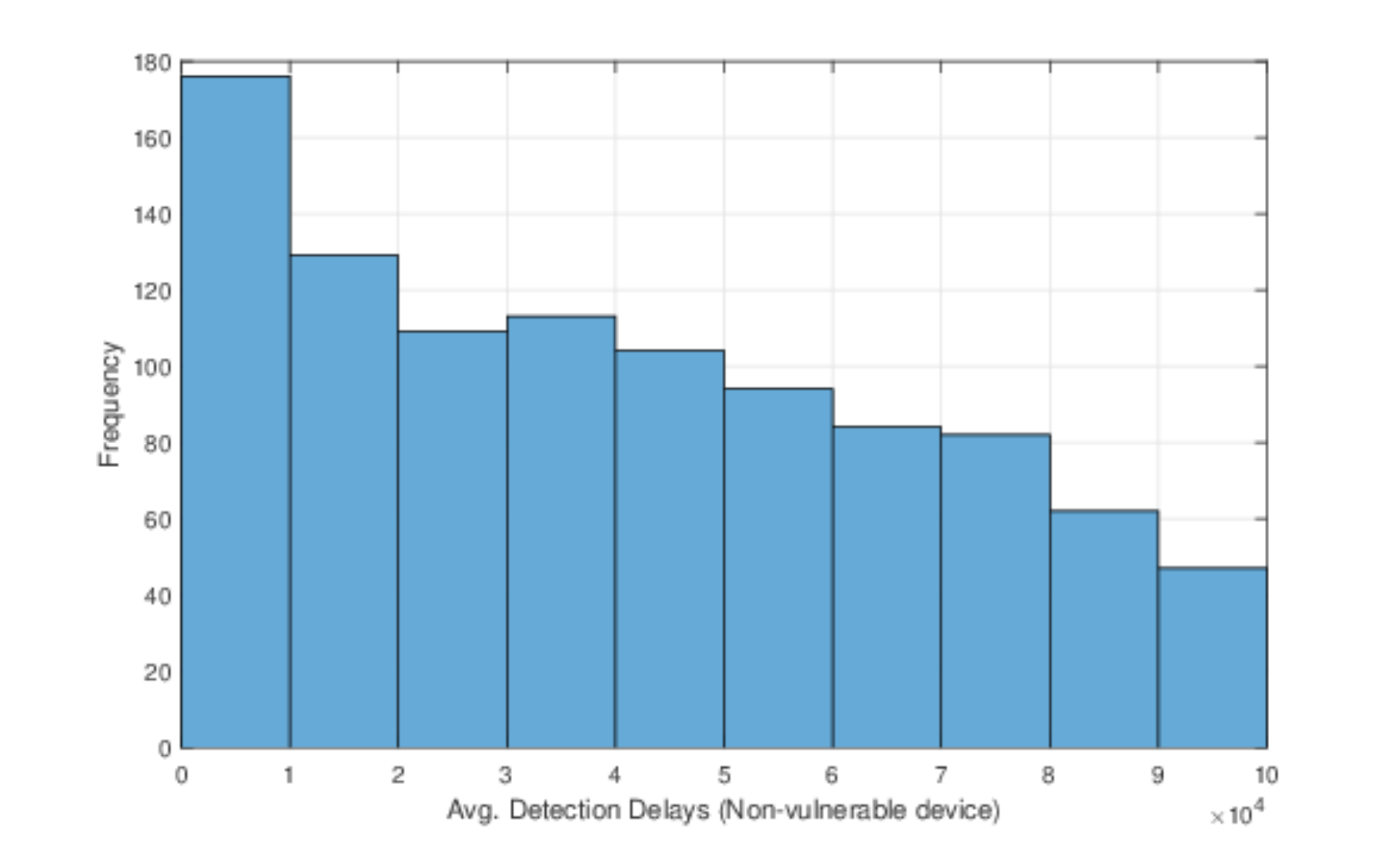}
\caption{Histogram of Average detection delays for non-vulnerable devices}
\label{sim4}
\end{figure}

In Fig. \ref{sim3}, we have presented the plot for average detection delay vs sampling frequency for different values of attack probability on non-vulnerable devices ($p_2$). The plot behavior is somewhat irregular near lower sampling frequencies. For higher sampling frequencies, the average detection delay can be observed to decrease almost linearly with increasing sampling frequency. The intuitive explanation for the decreasing behavior is similar to the one given above for vulnerable devices. While designing the proposed Algorithm \ref{A1}, a sampling frequency for non-vulnerable devices which is too high may lead to lower average detection delay but the corresponding increase in processing power and memory requirements may not be desirable since non-vulnerable devices are not expected to be compromised easily. A sampling frequency which is too low on the other hand, may increase the average detection delay significantly in the unexpected scenario when some of the non-vulnerable devices are compromised. Therefore, the algorithm designers may have to settle for a sampling frequency which falls in the mid of the range of available values. Fig. \ref{sim4} shows the distribution of average detection delays for non-vulnerable devices for a sampling frequency of 0.025 and attack probability of 0.2 using a histogram. The distribution assumes the highest values for average detection delays between `0-10,000'. Thereafter, values taken by the distribution decrease slowly with increasing average detection delays.  

\section{Future Work}
We are developing a software prototype of the proposed bot detection algorithm \cite{algocode} which will be evaluated on a testbed consisting of physical IoT and non-IoT devices, gateways, and routers. The network behavior of Mirai will be emulated by replaying Mirai traffic captured from our virtualized testbed. 
In the future, we would like to develop solutions for detecting IoT bots infected with malware exploiting software vulnerabilities to hack the devices and add to the botnet. For instance, Linux.Darlloz, Reaper and Amnesia malware \cite{linuxdarlloz,reaper,amnesia} use HTTP (Hyper Text Transfer Protocol)-based exploits to perform code injection and arbitrarily execute code on remote devices bypassing authentication. 
It should be noted here that the packet sub-sampling approach proposed in this paper is likely to be a part of the bot detection solution devised for such advanced malware. 
Finally, some malware may try to evade detection, e.g. by attempting to hide their scanning activity. It would be an interesting problem to detect such evasive IoT malware.  

\section{Conclusion}
In this paper, we proposed an algorithm for detecting IoT devices infected by Mirai or similar malware. The bot detection algorithm uses Mirai traffic signatures and a two-dimensional sub-sampling approach. The deployment of our algorithm within a real-world network was discussed and prospective actions which can be taken after bot detection were also mentioned. Leveraging measurements taken from a testbed constructed to emulate the behavior of Mirai, we studied the relationship between average detection delays and sampling frequencies for vulnerable and non-vulnerable devices. Based on our analysis of the plots, we made suggestions regarding the process of selection of sampling frequencies while designing our proposed algorithm. Finally, we identified few interesting problems stemming out of this research which we would like to work upon in the future.    

\section*{Acknowledgment}
The authors would like to thank Dr. Liang Zhenkai (SoC, NUS) for helping us with some of the initial ideas used in this paper and Dr. Min Suk Kang (SoC, NUS) for providing comments on our manuscript. We would also like to appreciate the National Cybersecurity R\&D Lab, Singapore for allowing us to use their testbed to collect important data which has been used in our work. This research is supported by the National Research Foundation, Prime Minister’s Office, Singapore under its Corporate Laboratory@University Scheme, National University of Singapore, and Singapore Telecommunications Ltd.

\bibliographystyle{splncs}
\begingroup
\raggedright
\bibliography{oqeprop}
\endgroup

\end{document}